\def\Nf{N^{\varphi}}

\def\mmax{M_{\rm max}}
\def\mbcr{M_{\rm B,crust}}
\def\tcr{\Delta R_{\rm crust}}
\def\mb{M_{\rm B}}
\def\mbacc{M_{\rm B,acc}}
\def\msol{{\rm M}_{\odot}}
\def\nurot{\nu_{\rm rot}}

\def\nukep{\nu_{\rm K}}

\def\rms{R_{\rm ms}}
\def\xl{x_{l}}
\def\lms{l_{\rm ms}}
\def\ems{e_{\rm ms}}

\def\bb{Br}
\def\pf{p_{\varphi}}
\def\uc{U_{\rm s}}
\def\et{{\bf{e_t}}}
\def\ef{{\bf{e_{\varphi}}}}
\def\er{{\bf{e_r}}}
\def\eth{{\bf{e_{\theta}}}}
\def\etr{{\bf{e_{\widehat t}}}}
\def\efr{{\bf e_{\widehat\varphi}}}
\def\err{{\bf e_{\widehat r}}}
\def\ethr{{\bf e_{\widehat\theta}}}

\def\ptr{{p^{\widehat t}}}
\def\pfr{{p^{\widehat\varphi}}}
\def\prr{{p^{\widehat r}}}

\documentclass{aa}
\usepackage{graphicx}
\begin{document}
%
%
%
\newcommand{\ddp}[2]{\frac{\partial #1}{\partial #2}}
\newcommand{\ddps}[2]{\frac{\partial^2 #1}{\partial #2 ^2}}

\title{Recycling strange stars to millisecond periods}
\author{J. L. Zdunik \inst{1}
 \and
P. Haensel\inst{1,2}
 \and
E. Gourgoulhon\inst{2}}
\institute{N. Copernicus Astronomical Center, Polish
           Academy of Sciences, Bartycka 18, PL-00-716 Warszawa, Poland
{\em jlz@camk.edu.pl}} \institute{N. Copernicus Astronomical
Center, Polish
           Academy of Sciences, Bartycka 18, PL-00-716 Warszawa, Poland
\and D\'epartement d'Astrophysique Relativiste et de Cosmologie --
UMR 8629 du CNRS, Observatoire de Paris, F-92195 Meudon Cedex,
France\\
{\em jlz@camk.edu.pl,  haensel@camk.edu.pl,
  Eric.Gourgoulhon@obspm.fr}}
\offprints{J.L. Zdunik}
\date{Received 28 August 2001 / Accepted 24 October 2001}
\abstract{Recycling strange stars to millisecond periods is
studied within the framework of general relativity.
We employ equations of state of strange quark matter
based on the MIT Bag Model,  with massive strange quarks and lowest
order QCD interactions. The presence of the crust of normal matter is
taken into account, with a bottom density assumed to be equal to the
neutron-drip one. The calculations are performed by solving the exact
2-D equations for rigidly rotating stationary configurations
in general relativity. Evolutionary tracks of accreting strange
stars are computed, and their dependence on the initial strange star mass
and on the fraction of the angular momentum transferred to
the star from the infalling matter, is  studied. The differences
between recycling strange stars and neutron stars are pointed
out.
}

\maketitle


\section{Introduction}
%
The problem of the spin up of the neutron (or strange) stars is an
important element of   the studies of recycled pulsars.
It is widely believed that millisecond pulsars ($P<10$~ms)
are old neutron stars (age $\ga 10^9$~y),
 with low magnetic
field ($B\la 10^9$~G), which  had been spun up by the accretion
when they were members
 of the low-mass binaries (see, e.g.,
van den Heuvel \& Rappaport
\cite{hr92}). This evolutionary scenario of formation of millisecond
 pulsars recently obtained some observational support with the discovery
of the first  accretion-powered millisecond pulsar SAX J1808.4-3658
(Wijnands \& van der Klis 1998).

The detailed theoretical modeling of the spin-up scenario requires
the solution of several specific problems that involve
the knowledge of the neutron star
structure, that in turn depends on the equation of state
of neutron star matter.
In particular, one has to determine the accreted mass needed
to spin up a compact star to the observed  millisecond periods, within
a reasonable time, in low-mass X-ray binaries. Finally, one would like
to know what limits the maximum accreted mass - whether
the upper bound results
from the mass-shedding limit, or from
some instability against gravitational collapse.
The answers to all these questions depend on the (largely unknown)
equation of state of dense matter at supranuclear density.

Klu{\'z}niak \& Wagoner (\cite{KW85})
studied the spin up of an accreting neutron star (NS)
for several equations of state of dense matter,
using the lowest-order expansions
in stellar angular momentum (slow rotation
approximation, pioneered by Hartle 1967).
Cook, Shapiro \& Teukolsky (\cite{CST94b}, hereafter CST)
modeled the recycling of NS without
any restriction on neutron-star rotation,
using their previous results for stationary rigidly
rotating stellar configurations, which were obtained
in the framework of general relativity
for fourteen  equations of state of dense matter
(Cook et al. \cite{CST94c}). A semianalytical model,  based on the
CST results, has been used by Burderi et al. (\cite{Burderi99}) to
study the dependence of the recycling  scenario on the magnetic field.

A hypothesis that the true ground state of matter is a deconfined
self-bound quark plasma (Bodmer 1971, Witten 1984),  called
strange matter, implies the possibility of the existence of strange quark
stars (Witten 1984; first detailed models in Haensel et al. 1986,
Alcock et al. 1986). Recently, it has been argued that the accreting
compact objects in some low-mass X-ray binaries could be strange stars (SS),
and not NS (Bombaci 1997, Li et al. 1999a,b).

Previous studies
of the properties of rotating strange stars pointed out significant,
and in some cases strong and even qualitative
 differences between rapidly rotating
SS and NS. In particular, the difference in the distribution
of matter in the interior of
the rapidly rotating NS and SS was shown to imply a
significant difference in the outer spacetime. Specifically,
qualitative differences in the properties of the innermost stable
circular orbit (ISCO)  around these two
classes of rapidly rotating compact
objects have been found (Stergioulas et al. 1999;
Zdunik et al. 2000, Zdunik \& Gourgoulhon \cite{ZG01}).
As  shown in Zdunik et al. (2001, hereafter Paper I),
a careful treatment of the crust of SS
to some extent softens these
conclusions,  but important differences still remain.
Recently,  parameters of marginally stable orbits and
properties of the accretion disk around SS have been studied,
and compared with those around NS,  by
Bhattacharyya et al. (\cite{BTB01}).

Differences between rapid rotation of SS and NS, as well as
those between properties of the space-time around them,
indicate the need for a study of differences between
recycling by accretion of SS and NS. In the present paper
we study spin-up of SS by accretion of matter from the innermost
stable circular orbit. We neglect the effect of
the magnetic field, accretion, and radiation drag on the
location of the ISCO, and on the infall of accreting matter,
which is justified for  $B\la 10^9~$G and $\dot{M}\ll
\dot{M}_{\rm Edd}$.
 Our approach is thus similar to that of CST. We have recently
calculated the models of
rotating strange stars with crust in general relativity
(Paper I).
Using the method described in Paper I, we determine the
parameters necessary to calculate the motion of the matter
accreting from the ISCO in the equatorial plane. The equation of state
of strange matter, as well as the method for solving the relevant
equations in general relativity, are briefly described in the
first part of Sect. 2. The remaining part of Sect. 2 is devoted
to the formulation of the basic equations governing the evolution of
an accreting SS. Numerical results of computation of the evolutionary
tracts of accreting SS are presented in Sect. 3, where we also
study the basic differences between spin-up by accretion of SS and  NS.
Finally, Sect. 4 is devoted to the discussion of our results and
gives some conclusions.
\section{Calculations}
The equation of state and calculational method for the stellar model
are the same as
that used  in Paper I.
Our EOS of strange matter, composed of massless u, d quarks, and
massive s quarks, is based on the MIT Bag Model with the standard values of
the Bag Model parameters for strange matter:  bag constant $B=56~{\rm
MeV/fm^3}$, mass of the strange quark $m_{\rm s}=200~{\rm MeV/c^2}$,
and QCD coupling constant $\alpha_{\rm c}=0.2$
(Farhi \& Jaffe 1984, Haensel et al. 1986, Alcock et al.
1986).   This EOS of strange quark matter (called SQM1 in Paper I)
had been also used in Zdunik et al. (2000) and
Zdunik \& Gourgoulhon (\cite{ZG01}).
It yields an energy per unit baryon
number at zero pressure $E_0=918.8 ~{\rm MeV} <E(^{56}{\rm
Fe})=930.4~$MeV. The maximum allowable mass for
static strange stars is $M_{\rm max}^{\rm stat}=1.8~{\rm M_\odot}$.
The crust in our model is described by the BPS EOS
below the neutron drip (Baym et al. 1970).

The general relativistic models of stationary rotating strange
stars have been calculated by means of the code developed by
Gourgoulhon et al. (1999), which relies on the multi-domain
spectral method introduced in Bonazzola et al. (1998).
In the present calculation  we use 4 spatial
domains (for details see Paper I).
We supplemented the version of the code used in Paper I
with the equations determining the  specific angular momentum
$l_{\rm ms}$ and specific energy $e_{\rm ms}$ of a test particle
falling from the ISCO to the surface of the star (see the Appendix).
We determine also the
angle $\beta$,  at which the particle falling freely from
the ISCO hits the surface of the star,
\begin{equation}
\tan(\beta) ={\prr\over \pfr}
\label{angle}
\end{equation}
where $p^{\hat{r}}$, $p^{\hat{\varphi}}$ are radial and tangential
momentum of the infalling test particle,
in the frame comoving with the matter at the stellar
surface. For energetic considerations (heating of
the surface), as well as for determining the probability of crossing
the electrostatic barrier at the surface of SS,
the crucial parameter is also the
energy of the particle as seen by the observer comoving with the star.
This energy is defined by the time component of the particle
4-momentum: $E=\ptr$.
The equations defining the parameters of the particle hitting the
surface of the star are presented in the Appendix.

The evolutionary tracks of the accreting SS are calculated using
the simplest scenario in which the increase of the angular
momentum of the star is related to the  accreted rest mass
increment by
\begin{equation}
\delta J = \xl\, \lms\, \delta \mb~,
\label{djdm}
\end{equation}
where we introduced the parameter $\xl$ - the fraction of the total
angular momentum of the particle transferred to the star. In the
idealized, limiting  case the accreted mass element transfers
all its angular momentum to  the star ($\xl=1$).
Equation (\ref{djdm}) allows us to determine the
 total angular momentum  $J$ and and total rest (baryon) mass
$\mb$, after accreting matter of rest mass
$\Delta \mb$:
\begin{eqnarray}
\mb&=&M_{\rm B,0}+\Delta \mb \label{evolm}~,\\
J&=&J_0+\int_0^{\Delta \mb} \xl\,\lms (\mb,J)\, {\rm d}\mb~,
\label{evolj}
\end{eqnarray}
where $M_{\rm B0}$ and $J_0$ are initial baryon mass and angular momentum
of the star, respectively. The values of $M_{\rm B}$ and $J$ determine then
in a unique way the instantaneous configurations of accreting star, which
form a one parameter family  of rigidly rotating configurations, labeled
by $M_{\rm B}$.
 These configurations form evolutionary tracks of accreting SS.
 The evolutionary tracks will be  labeled
by the gravitational mass $M_0$ of the  initial configuration
with  baryon mass
$M_{\rm B,0}$. This  initial configuration is assumed to be
nonrotating,
 $J_0=0$.

For the star accreting at the rate $\dot\mb$
(as seen by a distant observer),
the time evolution of $J$ is given by the equation
\begin{equation}
{{\rm d}J\over {\rm d}t} = \xl\, \lms\, \dot{\mb}~,
\label{djdt}
\end{equation}
resulting directly from Eq. (\ref{djdm}), where $t$ is the
time measured at infinity.

\section{Results}
\subsection{Changes of
$\lms$, $\ems$ along the evolutionary tracks}
As one can see from Eq.(\ref{djdt}), the quantity determining the
evolution of an accreting star is the specific angular momentum of the particle
at the ISCO: $\lms$. In the case of spherical,  nonrotating configuration
 with $R<\rms$,  the value obtained
for the Schwarzschild metric is $\sqrt{12}\, GM/c$.
For a rotating star, inclusion of the lowest order
correction in a dimensionless parameter
$j=Jc/GM^2$ leads to the formula  derived by Klu{\'z}niak
\& Wagoner (\cite{KW85}): $\sqrt{12}(1-{1\over3}\sqrt{2\over3}\,j)\,GM/c$.
Higher-order  expansions  in $j$ can be found in Shibata \&
Sasaki (\cite{SS98}) (Eq.~(B2) of their paper, where $j$ is denoted by $q$).
In this article, we use instead the exact formula for $\lms$,
as given in the Appendix.

\begin{figure}
\resizebox{\hsize}{!}{\includegraphics{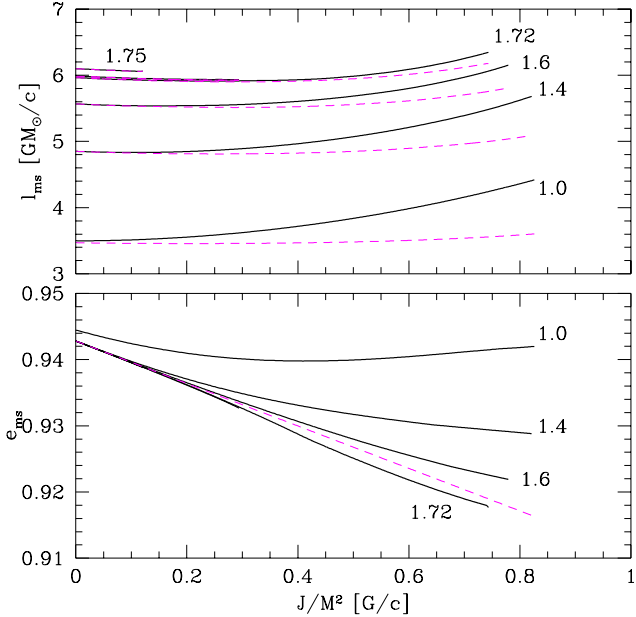}}
\caption{Angular momentum and energy (per unit mass)
of a particle at the marginally stable orbit along
an evolutionary track for
initial gravitational masses
$1, 1.4,1.6, 1.72, 1.73, 1.75 \,\msol$
and linear
approximation in  $j=Jc/GM^2$ of Klu{\'z}niak and Wagoner (1985)(dashed).
In the case of energy the curves for $1.73, 1.75 \,\msol$ cannot be
graphically distinguished from the upper part of the $1.72 \,\msol$
curve. The values of $\lms$
and $\ems$ for nonrotating configuration are
$\sqrt{12}\,GM/c$ and $2\sqrt{2}/3$,  respectively
(except for the $1\,\msol$ case, where the ISCO is defined
by the stellar radius).}
\label{lemsj}
\end{figure}

In Fig. \ref{lemsj} we present the value of the angular momentum and energy
of the test particle at the
ISCO along evolutionary tracks. The approximate formula
underestimates the value of $\lms$ and the actual spin up of the star
is larger than predicted by the linear approximation (in  $j$) of
Klu{\'z}niak \& Wagoner (1985). The energy of
the test particle at the ISCO,  $\ems$,  is smaller
than that given by the linear approximation in
$j$  for high masses only (the initial mass $M_0>1.7~\msol$).
This is due to the fact that the next order
terms in the expansion of $\ems$ and
$\lms$ in  $j$ are negative, while simultaneously
 rotational  deformation of the star leads to the
increase of both
$\ems$ and $\lms$. The negative correction, due to $j$,
dominates only for the most
 massive stars,  when the deformation due to the
rotation is relatively small because of the very strong  gravitational pull.
Such a cancellation is not so pronounced for $\ems$,
because there the relative magnitude of the
$j^2$ correction is about twice that for $\lms$
(see Shibata \& Sasaki \cite{SS98}, Eqs.~(B1,B2)).

\begin{figure}
\resizebox{\hsize}{!}{\includegraphics[angle=-90]{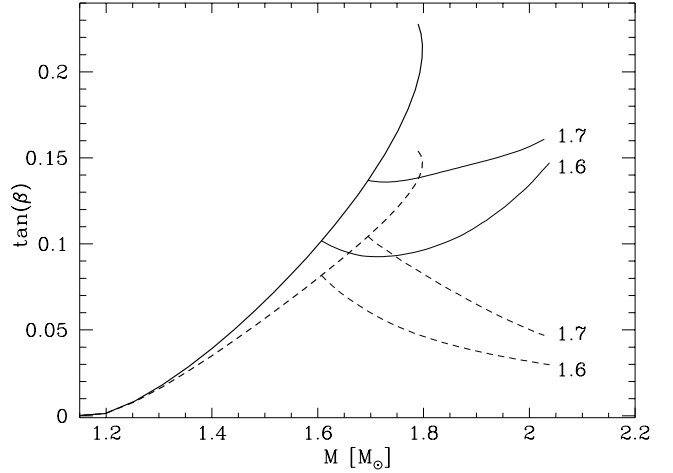}}
\caption{The ratio $p^{\hat{r}}/p^{\hat{\varphi}}$,
 equal to the tangent of
the incidence (impact) angle $\beta$,
 at which a test particle falling freely from the marginally
stable orbit hits the stellar surface.
Thick solid line - exact results for nonrotating
configurations up to the maximum mass. Thin solid line -
 evolutionary tracks for
$M_0=1.6\msol$ and $M_0=1.7\msol$. Dashed lines - same quantities,
calculated using the approximate formula
(Klu{\'z}niak \& Wagoner \cite{KW85}). Evolutionary tracks
terminate at the mass-shedding limit.}
 \label{angm}
\end{figure}

Figure \ref{angm} shows the value of the tangent of the
incidence (hitting) angle of the accreted
matter,   falling freely from the marginally stable orbit, as
measured in the rest frame comoving with  the stellar surface.
The deviation
of the approximate formula of Klu{\'z}niak and Wagoner (\cite{KW85})
from the exact results is quite large,
 reaching 50\% at maximum mass
of the nonrotating configurations, and is even greater
 for rapidly rotating
massive stars.
The main reason for  this discrepancy is the large gap
between the stellar surface and marginally stable orbit for strange stars.
For instance, at
 $M=\mmax$ and $\nurot=0$  we have $R\simeq 10$~km and
$\rms\simeq 16$~km, so that the linear  approximation of
the expansion of $\tan {\beta}$ in
$(\rms-R)/R$,  used by Klu{\'z}niak \& Wagoner (\cite{KW85}),  is
unjustified.

\begin{figure}
\resizebox{\hsize}{!}{\includegraphics[angle=-90]{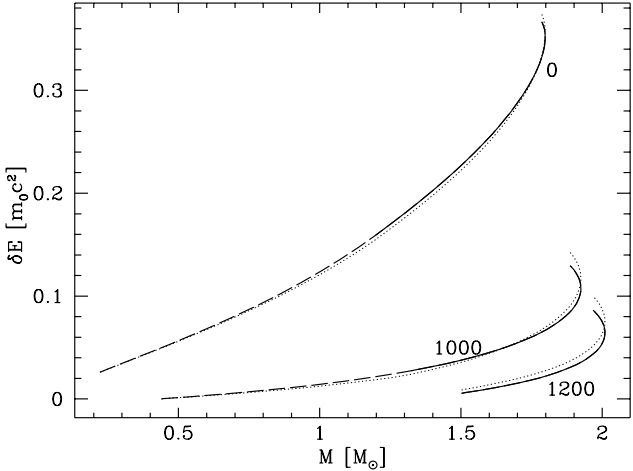}}
\caption{Kinetic energy
 of a test particle falling freely from the marginally
stable orbit as it hits the stellar surface (in the frame
comoving with the surface of the star).
The curves correspond to fixed rotational frequencies:
$\nurot =0,\,1000,\,1200$~Hz.
Thick lines - exact results (dashed part - ISCO defined by the stellar
radius). Thin dotted lines - nonrelativistic approximation
with relativistic time correction (see text).}
 \label{enm}
\end{figure}

In Figure \ref{enm} we present the value of the
energy of the infalling particle (minus its rest mass $m_0c^2$)
as seen by the observer at the
stellar surface.
This energy can be interpreted as the kinetic
energy of the particle which should be lost as it
gets a part of the star.
As one can expect the energy tends to 0 as we are approaching Keplerian
limit with $R>\rms$ (left part of $\nurot=1000$~Hz curve), since then
the Keplerian velocity at the equator is equal to the velocity of the
rotation.
The exact results are compared with the
approximate Newtonian formula (with a correction involving the relativistic
time delay which changes the velocities as seen by the observer
at the massive star):
\begin{equation}
\delta E = {1\over 2}{(v_{\rm ms}-v)^2\over(1-{2GM/
Rc^2})}+{GM\over R} \left( 1-{R\over \rms} \right) \ ,
\label{enewt}
\end{equation}
where $v=\Omega R$ is the velocity of the star at the equator and
$v_{\rm ms}=\Omega_{\rm ms} R_{\rm ms}$ is the velocity of the
particle at the marginally stable orbit.
This formula approximates $\delta E$ with a surprisingly small error.

\subsection{Accreted crust}
Standard models of strange matter predict the existence of an
electron layer, which extends beyond the sharp quark surface
by some $\sim 300-400~$fm. The electrons in this layer are
bound to the quark surface by a positive Coulomb potential,
produced by quarks, of height $\sim 30-50$ MeV. Spherical accretion
of a plasma onto a SS with low  magnetic field ($B\la 10^9~$G)
was studied in (Haensel \& Zdunik 1991, Miralda-Escud{\'e} et al.
1990). In the case of spherically symmetric accretion,
protons at the instant of hitting a bare (no crust)  SS surface
have an energy many times larger than the height of the Coulomb
barrier, and traverse it with a large probability. After being
absorbed by the strange matter, the protons (nucleons) dissolve into
quarks in an  exothermic process.

As we see in Fig. \ref{angm}, in the case of accretion from the
ISCO, nuclei in the infalling plasma  hit the SS surface at a very
oblique angle $\beta\la 10^{\rm o}$; they should therefore be
efficiently deflected by the electrostatic barrier. The kinetic energy
of a proton hitting the SS surface can be estimated as $0-300$~MeV
(Fig. \ref{enm}), and depends very sensitively on the value of
$\Omega$. The largest value of $\delta E$ corresponds to the
nonrotating star with maximum mass, since then the two components
defining this energy are very large: the velocity of the freely
falling particle relative to the nonrotating star (the kinetic
energy) and the gap between the marginally stable orbit and
stellar surface (the potential energy).
However a very small incidence angle increases strongly the effective
thickness of the
Coulomb barrier to be penetrated, as well as the effective
thickness of the
electron layer  to be traversed (implying therefore
higher energy loss). All in all, infalling
 protons (or more generally
- nuclei) are not able to penetrate the Coulomb barrier, and one
expects a gradual building-up of a normal-matter envelope
(crust) on the top of the strange matter surface. The bottom density
of the accreted crust is limited by the neutron-drip. Further
accretion induces irreversible absorption of neutrons by strange
matter, and their fusion with strange matter in a strong-interaction
process $n\longrightarrow
2d + u$, followed by the weak-interaction process
$u+d\longrightarrow s+u$. These reactions lead to an additional
heat release (deconfinement heating) at the crust-strange matter
interface, of some $30-40~$MeV per accreted nucleon.
%

\subsection{SS parameters along evolutionary tracks}

\begin{figure}
\resizebox{\hsize}{!}{\includegraphics[angle=-90]{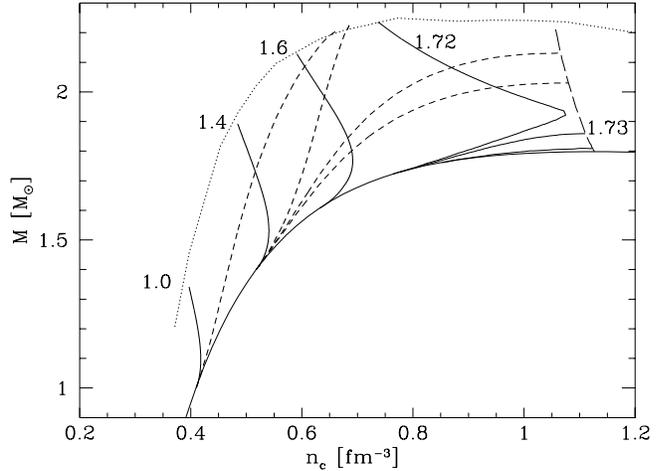}}
\caption{Gravitational mass versus central baryon number
 density along several
evolutionary tracks of accreting strange stars. Solid lines
correspond to the typical case - all angular momentum of
matter falling from the marginally stable orbit is
transferred to the star ($\xl=1$). Dashed lines
represent  three cases for a
$M_0=1.4~\msol$ star ($\xl=0.66,0.55,0.5$ from
the left to the right),  and  the case
 $\xl=0.5$ for a $M_0=1~\msol$ star. Long-dashed,
nearly vertical line at maximum mass is defined by the onset of
 instability with respect to axisymmetric perturbations
(see the text), dotted line - Keplerian (mass shedding) limit.}
 \label{mnc}
\end{figure}

In Fig. \ref{mnc} we plot  the gravitational
mass of the star as a
function of the central density,  for different choices of
$M_{\rm B,0}$ and for $J_0=0$. This figure can be directly compared
with Fig. 1 of CST, obtained for NS.
The main differences between neutron and strange stars spun up by
the accretion are the following:
\begin{itemize}
\item the difference between maximum masses
for  rotating and nonrotating SS  is more than
$0.4~\msol$ compared with at most $0.3~\msol$ for NS;
\item the specific angular momentum
$\lms$ for SS is sufficiently high to result in a
characteristic ``turning point''
of the evolutionary track
 in the $\rho_c - M$ plane,  at which
the central density of accreting SS starts to decrease.
The effect of the increasing
angular momentum (leading to the deformation of the star and
decrease of $\rho_c$) starts there to be
dominating over the the effect of the mass
increase tending to increase the central density.
\item the rapid increase of $J$, compared to NS,
determines the final fate of the accreting SS:
nearly all evolutionary sequences terminate at the mass-shedding
limit. Instability  with respect to axisymmetric perturbations
(which would eventually lead to the collapse of accreting SS into a
black hole) can be reached only for initial (static) mass
 very close to the maximum one for $x_l=1$; for our EOS this is restricted
to  the narrow range $1.73~\msol<M_0<\mmax=1.8~\msol$.
\item  in the case of accreting SS, there exists a region of stable
rotating configurations with high $M$ and high $\rho_{\rm c}$,
which cannot be reached via accretion of  matter onto
an initially  static  SS (unless $\xl<1$).
\item contrary to the case of NS, for astrophysically interesting
masses (say, $M>\msol$), we have,
 during accretion,  $R_{\rm eq}<\rms$.
This means, that along the evolutionary track
 the marginally stable orbit is always located above the stellar
surface. This conclusion is different from that  presented in Paper I,
because  it refers here
to the supramassive configurations, which have
not been considered in our previous paper.
\end{itemize}

\begin{figure}
\resizebox{\hsize}{!}{\includegraphics[angle=-90]{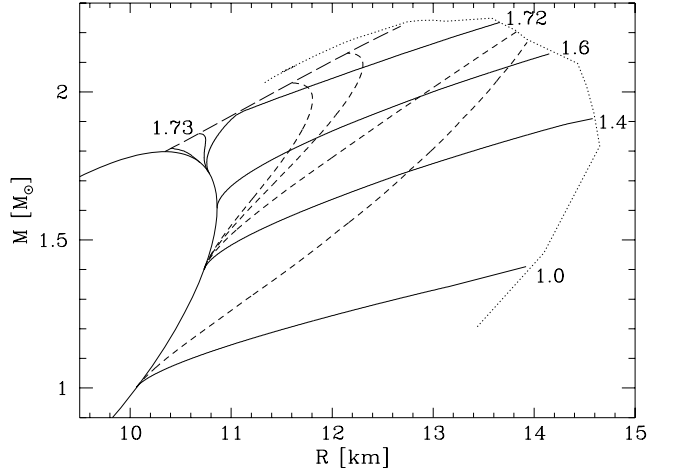}}
\caption{Mass versus the (circumferential) equatorial radius,
along  several evolutionary tracks of accreting strange stars.
Notations as in Fig \ref{mnc}.}
 \label{mr}
\end{figure}
The  mass-radius relations  for accreting SS
along evolutionary tracks are shown
in  Fig. \ref{mr}. The region that cannot be reached by the accretion
is not so spectacular as in the case of $n_{\rm c}-M$ dependence.
It is clearly visible that stars which, due to the accretion, directly enter
the region of instability with respect to axisymmetric perturbations
(implying collapse to black hole)
lie in a very narrow region ($1.9> M/\msol >1.73$, $11~{\rm km} > R
> 10.2~{\rm km}$). However it should be noted that in principle it
is possible that stars spun up by the accretion to the rotational
frequencies very close to the Keplerian one eventually are slowing
down due to some processes leading to the angular momentum loss
(violating of course Eq. (\ref{djdm}))
and reach the region forbidden from the point of view of
accretion.

\begin{figure}
\resizebox{\hsize}{!}{\includegraphics[angle=-90]{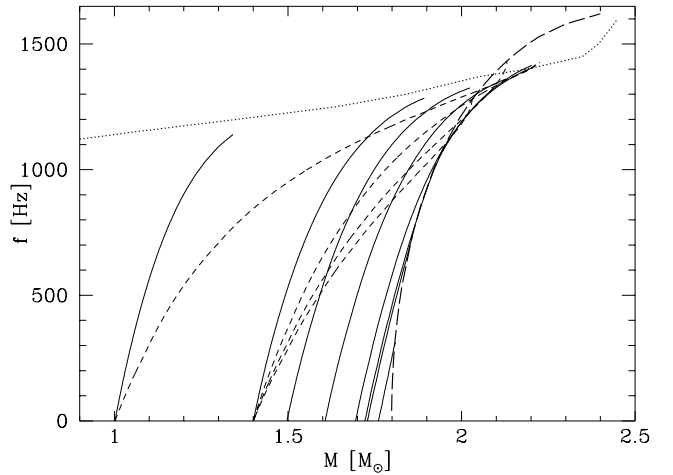}}
\caption{Frequency of the rotation versus gravitational mass 
along  several evolutionary tracks of accreting strange stars.
Notations as in Fig \ref{mnc}.}
 \label{fm}
\end{figure}
In Fig. \ref{fm} we present rotational frequency and gravitational mass
of the star - two quantities which could in principle be observed.
It should be stressed that the point in the $f-M$ plane do not  
uniquely defines the rotating configuration. In particular the crossing point
between the Keplerian (dotted) and the instability limit (thick dashed) curves
at about $2.5~\msol$, 1400 Hz do not correspond to the same star 
(the Keplerian limit star has a total angular momentum 
$J=4.2~G\msol^2/c$ and central density $n_c\simeq 0.5~{\rm fm}^{-3}$
 while the marginally stable star has $J=2.6~G\msol^2/c$ and 
$n_c=1.1~{\rm fm}^{-3}$).
As a consequence the region below the instability line and above the
Keplerian limit (dotted) curve for supramassive stars cannot be treated as 
forbidden (stable rotating configurations exist there)
although it can be reached by the accretion only in the
case of $\xl<1$.
\begin{figure}
\resizebox{\hsize}{!}{\includegraphics[angle=-90]{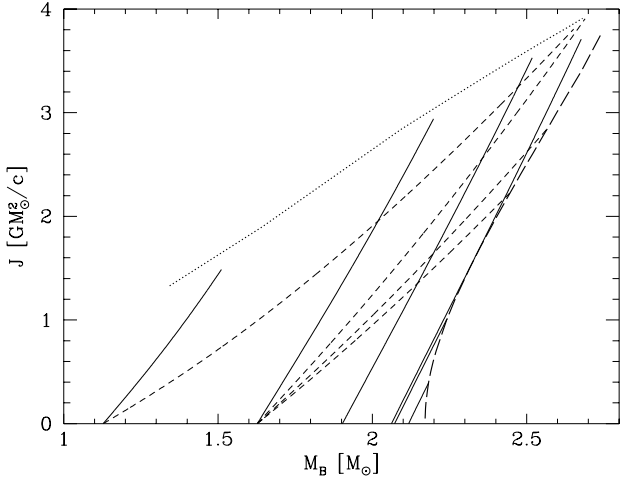}}
\caption{Angular momentum of the star versus its baryon mass,
along several evolutionary tracks of accreting strange stars.
At each point, the  slope (first
derivative) of evolutionary  curves is equal to the value
of $\xl \lms$. Notations as in Fig. \ref{mnc}}
 \label{jmb}
\end{figure}

The existence of the region which cannot be reached by the
accretion can be suitably discussed in  the $J-\mb$ plane -
two basic parameters from the point of view of accretion
(Fig. \ref{jmb}).
The evolutionary tracks are defined by the requirement that the
change of the stellar angular momentum is connected with
the accreted mass via Eq.~(\ref{djdm}), i.e., the derivative
${\rm d}J/{\rm d} \mb$ along evolutionary tracks is equal to
$\xl \lms$. The allowed region is bounded by the Keplerian
limit (high $J$) and by the instability with respect to
axisymmetric perturbations (high $\mb$) leading to the collapse.
As it can be seen from Fig. \ref{jmb} the accreting star eventually
collapses if the evolutionary track crosses the instability limit
(long dashed curve in Fig. \ref{jmb}), which is possible if
${\rm d}J / {\rm d} \mb$ along this limit is larger than
$({{\rm d}J / {\rm d} \mb})_{\rm evol}$ along the evolutionary track.
For $\xl=1$ this condition limits the possible parameters of
collapsing accreting SS to a very narrow range (high $\mb$ and low
$J$) whereas for a smaller $\xl$ all configurations on the instability curve
can be reached by accretion. It should be noted that, contrary to Figs.
\ref{mnc} and \ref{mr}, Fig.~\ref{jmb} presents stellar configurations
in a non-unique way, i.e., above the long-dashed curve there exist stable
(lower $n_{\rm c}$, higher $R$) as well as unstable (higher $n_{\rm c}$,
 lower $R$) configurations. 
Stars below this curve do not exist, which is the
direct consequence of this stability criterion (minimum $J$ at
fixed $\mb$).

\begin{figure}
\resizebox{\hsize}{!}{\includegraphics[angle=-90]{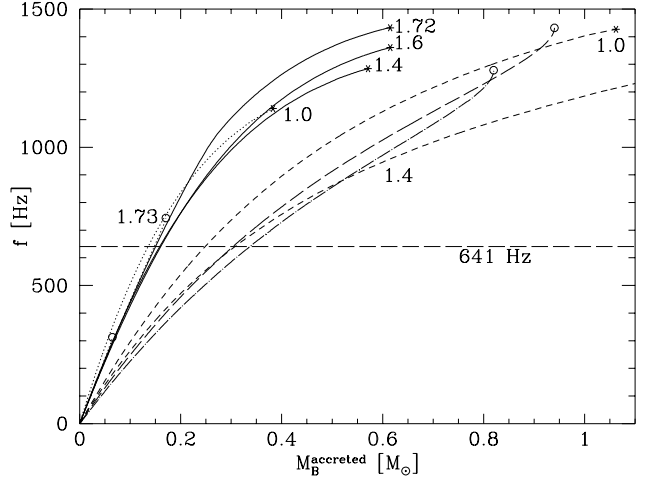}}
\caption{Rotational frequency of rotating strange star as a
function of the accreted baryon mass.  Dotted lines - 
$M_0=1, 1.73, 1.75~\msol$. Dashed lines
correspond to $\xl=0.5$ for $M_0=1~\msol$ and
$\xl=0.5$ (dashed), $\xl=0.55$ (dot-dashed), $\xl=0.66$ (long dash)
for $M_0=1.4~\msol$. The last points of the each evolutionary
track are labelled by the initial gravitational mass and denoted by a star
(mass-shed limit) or circles (collapse).}
 \label{fdmd}
\end{figure}

Figure \ref{fdmd} shows the rotational frequency
of a spun-up SS versus
the accreted  baryon mass.
 The evolution of the rotational period depends critically on the
 fraction  of the angular momentum at the ISCO,
transferred to the star, $\xl$. On the contrary, the character of the
spin-up depends
 rather weakly  on the initial mass of the star.
For  $\xl=1$,  the submillisecond periods are reached after accretion
of about $0.3~\msol$. Accretion of $0.2~\msol$ is needed
to  spin up the SS  to $1.56$~ms
($f=641$~Hz), the minimum observed period of a millisecond
pulsar.
Notice, that at a fixed $\dot\mb$, the  abscissa
in Fig.\ref{fdmd} is proportional to the accretion
time,  $t_{\rm acc}=M_{\rm B,acc}/\dot\mb$.

An interesting quantity, in the astrophysical context,
is the accreted rest mass required to spin up
the SS  to  the termination point of  the evolutionary track
(i.e., to the mass-shedding  limit,  or to the
threshold for instability with respect to the
axisymmetric perturbations, which results in collapse into a black hole).
 For the  evolutionary tracks  which eventually lead to
collapse into a black hole, this maximum accreted mass
depends sensitively on the initial mass $M_0$, i.e.,
 how close $M_0$ is to the maximum allowable mass.
 For example,  for
$M_0=1.75~\msol$, the  maximum accreted mass is only
$\mbacc=0.07~\msol$, while for
the lowest $M_0$, for which the evolutionary track terminates by
instability with respect to axially symmetric perturbations,
$M_0=1.723~\msol$, the SS accretes  $\mbacc=0.2~\msol$ before
collapsing into a black hole.
The maximum accreted mass,  required to spin up a SS
from $\nurot=0$ to
the mass-shedding limit $\nurot=\nukep$,  practically does not
depend on the initial mass
(provided it is not too small,  $M_0>1~\msol$),
and  is roughly equal $\mbacc=0.6~\msol$, i.e.,  is
larger by some 50\%  than in the case of neutron stars (see CST).
\begin{figure}
\resizebox{\hsize}{!}{\includegraphics[angle=-90]{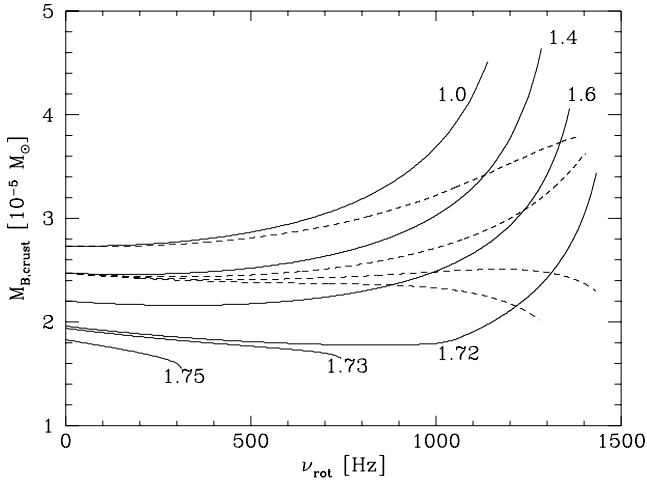}}
\caption{Baryon mass of the crust along evolutionary tracks.
$M_0=1,1.4,1.6,1.72,1.73,1.75\msol$.
Higher initial mass corresponds to lower $\mbcr$.
Dashed curves correspond to $\xl=0.5$ for $M_0=1\msol$ and
$\xl=0.5,0.55,0.66$ (from bottom to the top) for $M_0=1.4$}
 \label{mbcrf}
\end{figure}

\begin{figure}
\resizebox{\hsize}{!}{\includegraphics[angle=-90]{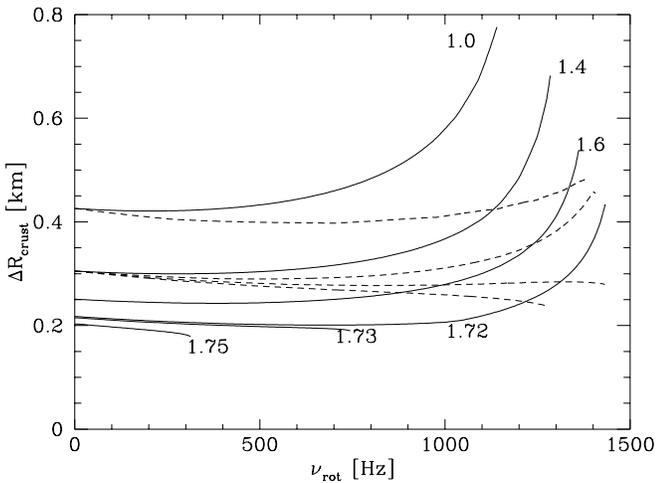}}
\caption{Thickness of the crust along evolutionary tracks.
Notations as in Fig. \ref{mbcrf}}
 \label{tcrf}
\end{figure}

The parameters of the crust (the baryon mass and the thickness) are
presented in Figs. \ref{mbcrf}, \ref{tcrf}. The interesting difference with
respect to the curves presented in Paper I
for a fixed baryon mass of the
star $\mb$ is the slower  increase (or even decrease)
of $\mbcr$ and $\tcr$ with increasing $\nurot$.
This is caused by the increase of the
SS mass during accretion: a larger mass
corresponds to higher gravitational pull and smaller crust (Paper I).
Although the rotation results in the increase of  $\mbcr$ and $\tcr$
at slow rotation rates the mass increase can
dominate,  leading to the shrinking  of the crust.
Of course,  the effect of the mass increase is more important,
if the
accreted mass is connected with a slower  acceleration of the rotation
(i.e. $\xl<1$). For example,  for $\xl=0.5$ the size of the crust of
an accreting $M_0=1.4\msol$ SS star is continuously
 decreasing along the
evolutionary track.

\section{Discussion   and conclusion}
Our study of SS recycling by accretion have shown
significant differences with respect to the
case of NS. These differences stem both from the
basic difference in the internal structure (density profile)
and from the difference in the outer space-time around
rotating SS and NS. The latter implies differences in the
properties of the marginally stable orbit, from which the matter
is falling onto the compact object, transferring its angular
momentum to the star.
Our calculations have been performed for a specific EOS of
strange matter. However, the qualitative features of evolutionary
tracks, described in Sect. 3, are generic for SS, and
do not depend  on the specific EOS of self-bound strange quark matter.

The standard models of strange matter (a recent review can be found in
Madsen 1999), predict the presence of a small fraction of electrons,
which is due to the SU(3) symmetry breaking in the quark sector
($m_s\gg m_u,m_d$). The electrostatic potential, which bounds
electrons to SS, produces then a huge outward-directed electric
field, which is necessary to support a crust of normal matter
on the strange matter core. Using such a   strange matter model,
one is  able to study the formation and evolution of the
crust during the accretion process.

Very recently, Rajagopal \& Wilczek (2001) demonstrated that
quark matter in the color-flavor locked (CFL) superconducting
phase of QCD is strictly neutral, despite unequal quark masses:
$n_u=n_d=n_s$. If strange matter is actually in the CFL
superconducting phase, then SS would be bare, with no
electrons, and with a surface thickness of $\sim \hbar c/\mu$, where
$\mu\sim 400~$MeV is the quark chemical potential. In absence of
any repulsive Coulomb barrier, the stable coexistence of normal and quark
matter would then be excluded. All accreted matter would be then
absorbed in an exothermic process. However, the effect on
the evolutionary tracks of accreting SS would be small.
The gap $R_{\rm ms}-R_{\rm eq}$ would be larger, and the
shedding-limit terminations would shift to somewhat higher frequencies,
allowing for accreting of a larger amount of matter.
However, the parameters of the marginally stable orbit, and the rate of
spin-up of accreting SS would not change, because  the
maximum fraction of stellar mass and moment of inertia contained
in a normal crust is at most $10^{-4}$.
\begin{acknowledgements}
This research
was partially supported by the KBN grant No. 5P03D.020.20
and by the  CNRS/PAN program  Jumelage Astrophysique.
\end{acknowledgements}

\appendix
\section{Free fall from marginally stable orbit - impact parameters}

In our considerations the metric can be written in the form:
\begin{eqnarray}
ds^2&=&-N^2dt^2+B^2r^2\sin^2\theta(d\varphi-\Nf
dt)^2\nonumber\\
&&+A^2(dr^2+r^2d\varphi^2)
\label{metric}
\end{eqnarray}
where $N$, $\Nf$, $A$ and $B$ are four metric functions of
$(r,\theta)$.
For a complete description of the coordinate system,
metric functions and relativistic equations see Bonazzola et al.
(\cite{bona93}) and Gourgoulhon et al. (\cite{Gour99}).
Here we present only the equations for the parameters of the
particle falling freely from the marginally stable orbit (ISCO).

The location (coordinate $r_{\rm ms}$) and orbital frequency
$\Omega_{\rm ms}$ of the ISCO are obtained by solving the
equations presented in e.g. Cook et al \cite{CST94a}, Datta et al.
1998, Schaab \& Weigel \cite{Schaab99}.
For any particle falling along a geodesic from the ISCO to the stellar surface,
the conserved quantities  are the specific energy
\begin{equation}
    e_{\rm ms} = - p_t = \Gamma_{\rm ms} (N + B r_{\rm ms} N^\varphi
                V_{\rm ms})     \label{e:e_ms}
\end{equation}
and the specific angular momentum
\begin{equation}
    l_{\rm ms} = p_\varphi = \Gamma_{\rm ms} B r_{\rm ms} V_{\rm ms} \ ,
                            \label{e:l_ms}
\end{equation}
where $p_\mu$ denotes the covariant 
components with respect to the coordinates
$(t,r,\theta,\varphi)$ of the particle 4-momentum (per unit mass),
$V_{\rm ms}$ is the particle 3-velocity as measured by the zero-angular-momentum
observer (ZAMO) and $\Gamma_{\rm ms}$ the corresponding Lorentz factor:
\begin{equation}
    V_{\rm ms} := {B \over N} r_{\rm ms}( \Omega_{\rm ms} - N^\varphi ) \ ,
\end{equation}
\begin{equation}
    \Gamma_{\rm ms} := \left( 1- V_{\rm ms}^2 \right) ^{-1/2} \ .
\end{equation}
In the above equation, all the metric factors are to be evaluated at the
location of the ISCO.

Let us now compute the impact angle and energy of the particle
as it reaches the stellar surface.
First we construct an orthonormal tetrad connected with the observer
comoving with surface of the star, the angular velocity of which is $\Omega$
(as seen from infinity). Let
$(\etr,~ \efr,~ \err,~ \ethr)$ denote this tetrad. It is expressed in
terms of the the coordinate basis $(\et,~ \ef,~\er, ~\eth)$
as follows:
\begin{eqnarray}
\etr&=&{\bf u}={\Gamma\over N}\et+\Omega {\Gamma\over N}\ef \nonumber\\
\efr&=&{\Gamma\over N}\uc\et + {\Gamma\over\bb}(1+{\bb\over
N}\Nf\uc)\ef \label{tetrad} \\
\err&=&{1\over A}\er\nonumber \\
\ethr&=&{1\over Ar}\eth \nonumber
\end{eqnarray}
where ${\bf u}$ is the 4-velocity of the stellar matter,
$\uc$ its 3-velocity with respect to the ZAMO,
\begin{equation}
\uc :={\bb\over N} (\Omega-\Nf) \ ,
\label{uc}
\end{equation}
and $\Gamma$ the corresponding Lorentz factor:
\begin{equation}
\Gamma := \left( 1-\uc^2 \right) ^{-1/2}\, .
\end{equation}

It is easy to check that this tetrad is orthonormal, i.e.
${\bf e}_{\widehat \mu}\cdot{\bf e}_{\widehat \nu}=
\eta_{\mu\nu}={\rm Diag}(-1,1,1,1)$.

The freely falling particle that hits the surface of the star
has a 4-momentum ${\bf p}$ with two conserved components:
the quantities $p_t=-\ems$ and $\pf=\lms$ given by Eqs.~(\ref{e:e_ms})
and (\ref{e:l_ms}). For a motion within the equatorial plane, the
component $p_\theta$ vanishes. The component $p_r$ is obtained from the
normalization relation ${\bf p} \cdot {\bf p} = -1$:
\begin{equation}
p_r^2=A^2\left[{(\ems-\Nf \lms)^2\over N^2}-{\lms^2\over
B^2r^2}-1\right] \ ,
\label{prd2}
\end{equation}
which of course is nonzero at the
surface of the star unless the ISCO is determined by
the stellar radius.

In the frame comoving with matter (tetrad \ref{tetrad}) the
momentum of the particle is given by:
\begin{eqnarray}
\ptr&=&-{\bf p}\cdot \etr=-(\pf (\etr)^{\varphi} +p_t
(\etr)^t)={\Gamma\over N} [\ems-\Omega \lms] \nonumber   \\
\prr&=&{\bf p}\cdot \err=p_r (\err)^r=p_r/A\nonumber \\
&=&\sqrt{{(\ems-\Nf \lms)^2\over N^2}-{\lms^2\over B^2r^2}-1}\\
\pfr&=&{\bf p}\cdot \efr=\pf (\efr)^{\varphi} +p_t
(\efr)^t\nonumber\\
&=&{\Gamma \lms\over \bb} \left[1+{\bb \uc\over N}
    \left(\Nf-{\ems\over \lms}\right)\right]\nonumber
\end{eqnarray}

The angle $\beta$ at which the particle falling freely from
the marginally stable orbit
 hits the surface of the star is given by
\begin{equation}
\tan\beta ={\prr\over\pfr} \ .
\label{angle1}
\end{equation}

The energy release per unit mass due of infalling matter in the comoving
frame is $\ptr-1$, i.e.
\begin{equation}
\delta E={\Gamma\over N} [\ems-\Omega \lms-N/\Gamma] \ .
\end{equation}
This is the value which should be used for the energy balance
 considerations
at the stellar surface (heating of the matter). If we assume that all of
this energy is emitted, this results in a luminosity (per unit accreted mass)
seen by a distant observer
${\delta E/ u^t}=\ems-\Omega \lms-N/\Gamma$,
as it has recently been demonstrated by Sibgatulin \& Sunyaev (\cite{SS00}).

\end{document}